\documentclass[]{spie}  
\usepackage{amsmath, amsfonts, graphicx, booktabs, tabularx, multirow, siunitx, xcolor}
\usepackage{hyperref}

\graphicspath{{./figures/}}

\title{Multiview and Multiclass Image Segmentation using Deep Learning in Fetal Echocardiography\thanks{\phantom{ee}This paper was accepted by SPIE Medical Imaging 2021. \copyright 2021 Society of Photo-Optical Instrumentation Engineers (SPIE). One print or electronic copy may be made for personal use only. Systematic reproduction and distribution, duplication of any material in this publication for a fee or for commercial purposes, and modification of the contents of the publication are prohibited.}
}

\author{Ken C. L. Wong$^1$, Ph.D., Elena S. Sinkovskaya$^2$, M.D., Alfred Z. Abuhamad$^2$, M.D., Tanveer~Syeda-Mahmood$^1$, Ph.D.
\skiplinehalf
$^1$IBM Research -- Almaden Research Center, San Jose, CA, USA\\
$^2$Eastern Virginia Medical School, Norfolk, VA, USA
}

\authorinfo{Please send correspondence to Ken C. L. Wong (clwong@us.ibm.com).}

\begin{document}
\maketitle

\begin{abstract}
Congenital heart disease (CHD) is the most common congenital abnormality associated with birth defects in the United States. Despite training efforts and substantial advancement in ultrasound technology over the past years, CHD remains an abnormality that is frequently missed during prenatal ultrasonography. Therefore, computer-aided detection of CHD can play a critical role in prenatal care by improving screening and diagnosis. Since many CHDs involve structural abnormalities, automatic segmentation of anatomical structures is an important step in the analysis of fetal echocardiograms. While existing methods mainly focus on the four-chamber view with a small number of structures, here we present a more comprehensive deep learning segmentation framework covering 14 anatomical structures in both three-vessel trachea and four-chamber views. Specifically, our framework enhances the V-Net with spatial dropout, group normalization, and deep supervision to train a segmentation model that can be applied on both views regardless of abnormalities. By identifying the pitfall of using the Dice loss when some labels are unavailable in some images, this framework integrates information from multiple views and is robust to missing structures due to anatomical anomalies, achieving an average Dice score of 79\%.
\end{abstract}



\section{INTRODUCTION}

Congenital heart disease (CHD) is the most common congenital abnormality leading to expensive hospital admissions associated with birth defects in the United States. Cardiac screening is currently performed by ultrasound examination of the fetus in the second and third trimesters of pregnancy. Despite training efforts and substantial advancement in ultrasound technology over the past years, CHD remains an abnormality that is most frequently missed by prenatal ultrasonography, with wide detection rates among various geographic regions and centers.

Computer-aided detection of CHD plays a critical role in prenatal care by improving screening and diagnosis. Since many CHDs involve structural abnormalities, automatic segmentation of anatomical structures is an important step in the analysis of fetal echocardiograms. Nevertheless, due to the low availability of labeled datasets in the public domain, automatic interpretation of fetal echocardiography has made limited progress. Most work has focused on segmentation in the four-chamber view (4CHV). Traditional techniques such as active appearance models \cite{Journal:Guo:TBME2014:Automatic} and region growing \cite{Conference:Punya:ICATIECE2019:Hybrid} have been used to segment the left and right ventricles. With deep learning approaches \cite{Journal:Xu:IEEEAccess2020:Convolutional,Conference:Yang:ISBI2020:Segmentation}, more anatomical structures can be segmented. However, due to the large annotation effort required, frequently the clinician demarkings have been restricted to chambers and aorta\cite{Conference:Yang:ISBI2020:Segmentation}. For a more complete characterization of the over 18 different types of structural heart defects such as tetralogy of Fallot, atrial septal defects, and ventricular septal defects \cite{Misc:webmdarticle}, a more thorough characterization of the heart regions would be needed. Fortunately, the national guidelines have recommended restricting to specific cardiac views such as the three-vessel trachea view (3VTV) and 4CHV for screening ultrasound examination. Even so, since echocardiography is sequence imaging, relevant keyframes have to be identified within which it is appropriate to segment the structures for interpretation. Such man-machine cooperation between AI algorithms and clinicians is ideal for commercial rendering of the technology as part of echocardiography systems to be used in clinical practice.

In this paper, we make two major contributions. First, we created a fetal echocardiographic dataset that represents a more detailed cataloging effort to map most meaningful structures in the heart for interpreting congenital anomalies. This dataset was developed by a team of physicians and sonographers with extensive expertise in fetal echocardiography at Eastern Virginia Medical School. Specifically, 14 structures in Table \ref{table:data} were considered sufficient for characterizing most CHDs in practice. These structures were annotated in the clinician-identified keyframes in 3VTV and 4CHV for optimal CHD interpretation. Next, we developed a deep learning framework for semantic segmentation of these 14 structures. We improve the V-Net \cite{Conference:Milletari:3DV2016} with spatial dropout, group normalization, and deep supervision, and the trained model can be applied on both 3VTV and 4CHV images of normal and abnormal cases. Despite the difficulties of missing structures, using the carefully adjusted exponential logarithmic loss function, we achieve an average Dice coefficient of 79\%.


\section{METHOD}

\subsection{Fetal echocardiograms}

A 2D fetal echocardiographic dataset with 199 normal and 100 abnormal cases was used. Two images per case demonstrating the 3VTV and 4CHV in 2D were retrieved from the cine-loops of the fetal heart stored in a prenatal imaging database. Each image was obtained at gestational age between 18+0 and 24+6 weeks and was selected during systole. Images with normal cardiac anatomy were selected if the anatomic landmarks specific for each view were clearly demonstrated. Anatomic landmarks for 4CHV include the presence of one complete rib on each side of the fetal lateral chest wall, visualization of the cardiac chambers, atrioventricular valves, descending aorta and fetal spine. Anatomic landmarks for 3VTV included visualization of pulmonary artery, aorta, superior vena cava, trachea and fetal spine. Cases with insufficient visualization of the above landmarks were excluded. Cases of major fetal CHD diagnosed prenatally with well documented 3VTV and 4CHV comprised the group of abnormal cardiac anatomy. All cases of CHD were confirmed by postnatal imaging, surgical findings and/or autopsy. Cardiac defects were considered to be major if surgery or intervention procedure was required during the first year of life. All images were fully de-identified before review. Each image was selected and manually segmented by physicians and sonographers with extensive expertise in fetal echocardiography, and this resulted in a total of 598 images with 14 anatomical labels (Table \ref{table:data}). Each image was zero-padded in the shorter side and resized to 1024$\times$1024. For some abnormal cases, some structures and thus the correspond labels do not exist.

\begin{table*}[t]
\caption{Semantic labels of anatomical structures and the associating views.}
\label{table:data}
\smallskip
\scriptsize
\centering
\begin{tabularx}{\linewidth}{XXXXX}
\toprule
\textbf{1. Left ventricle} & \textbf{2. Right ventricle} & \textbf{3. Left atrium} & \textbf{4. Right atrium} & \textbf{5. Descending aorta} \\
\midrule
4CHV & 4CHV & 4CHV & 4CHV & 4CHV \\
\toprule
\textbf{6. Pulmonary Artery} & \textbf{7. Aorta} & \textbf{8. Superior vena cava} & \textbf{9. Trachea} & \textbf{10. Spine}  \\
\midrule
3VTV & 3VTV & 3VTV & 3VTV & 3VTV, 4CHV \\
\toprule
\textbf{11. Interventricular septum} & \textbf{12. Interatrial septum} & \textbf{13. Mitral valve} & \textbf{14. Tricuspid valve} \\
\midrule
4CHV & 4CHV & 4CHV & 4CHV \\
\bottomrule
\end{tabularx}
\bigskip
\end{table*}

\subsection{Network architecture}

\begin{figure}[t]
    \centering
    \begin{minipage}[b]{0.95\linewidth}
      \centering
      \includegraphics[width=1\linewidth]{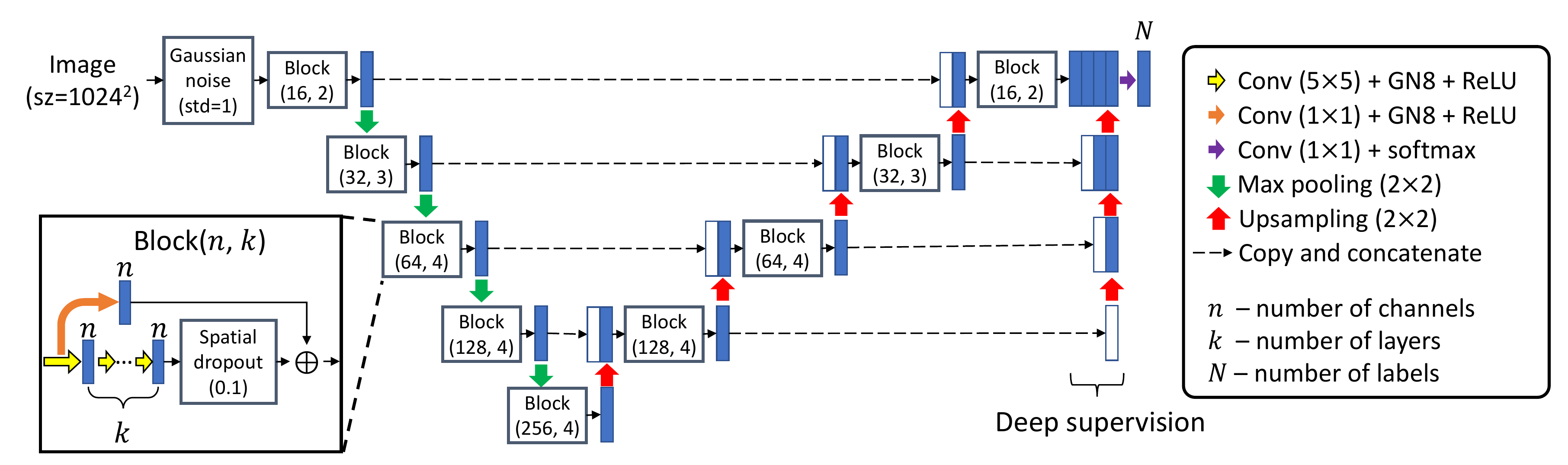}
    \end{minipage}
    \smallskip
    \caption{Network architecture. Blue and white boxes indicate operation outputs and copied data, respectively. GN8 stands for group normalization with eight groups of channels.}
    \label{fig:network}
\end{figure}

To segment the echocardiograms with relatively large image sizes, we modify the network architecture in \cite{Conference:Wong:MICCAI2018} which combines the advantages of low memory footprint from the V-Net \cite{Conference:Milletari:3DV2016} and the fast convergence from deep supervision \cite{Conference:Lee:AISTATS2015} (Fig. \ref{fig:network}). Each block in Fig. \ref{fig:network} comprises $k$ 5$\times$5 convolutional layers of $n$ channels, with spatial dropout \cite{Conference:Tompson:CVPR2015} and residual connection \cite{Conference:He:ECCV2016} to reduce overfitting and enhance convergence. The deep supervision part improves learning efficiency and thus convergence \cite{Conference:Lee:AISTATS2015,Journal:Dou:MedIA2017}. Given the image size of 1024$\times$1024, the batch size needs to be small (e.g., two) because of the memory requirements. As batch normalization does not work well with a small batch size, group normalization \cite{Conference:Wu:ECCV2018} whose performance is independent of the batch size is used, and eight groups of channels per layer gave the best performance in our experiments.

\subsection{Loss function}
\label{sec:loss}

As the object sizes of the labelled structures are highly unbalanced, the exponential logarithmic loss function in \cite{Conference:Wong:MICCAI2018} comprising the Dice loss and categorical cross-entropy is used. In our experiments, when only the data of a single view (e.g. 3VTV) were used, the loss function provided high accuracy. Nevertheless, when the data of both views were used together, for which not all labels exist in every image (Table \ref{table:data}), the validation Dice coefficients of some labels suddenly dropped to zero while the loss function was decreasing (i.e., improving). We found that the issue was caused by the argument $\epsilon$ in the Dice loss:
\begin{gather}
    L_\mathrm{Dice} = \mathbf{E}\left[\left( - \ln(Dice_l) \right)^{0.3}\right]
\end{gather}
with $l$ the label and $\mathbf{E}[\bullet]$ the mean value with respect to $l$. $Dice_l$ is the soft Dice coefficient:
\begin{gather}
\label{eq:soft_dice}
    Dice_l = \tfrac{2 \left(\sum_{i=1}^{N} p_{li} y_{li}\right) + \epsilon}{\left(\sum_{i=1}^{N} p_{li} + y_{li}\right)  + \epsilon}
\end{gather}
where $p_{li} \in [0, 1]$ are the network prediction scores, $y_{li} \in \{0, 1\}$ are the ground-truth annotations, and $N$ is the number of pixels in an image. $\epsilon$ is a small positive number (e.g., $10^{-7}$) to avoid the divide-by-zero situations. The use of $\epsilon$ does not cause any problem when all labels are available in every image. Nevertheless, if label $l$ is unavailable in an image (i.e., $y_{li} = 0$ for all $i$), we have:
\begin{gather}
    Dice_l = \tfrac{\epsilon}{\left(\sum_{i=1}^{N} p_{li}\right)  + \epsilon}
\end{gather}
for which the best $p_{li}$ are zeros as this gives $Dice_l = 1$. Therefore, depending on the number of images without label $l$, the optimizer may push all $p_{li}$ to 0 for all images to obtain the best overall loss. With this understanding, we remove $\epsilon$ from (\ref{eq:soft_dice}) and avoid the divide-by-zero situations by setting the lower bound of $p_{li}$ to a small positive number (e.g., $10^{-7}$). In this sense, when label $l$ is missing, $Dice_l$ is always 0 regardless of the values of $p_{li}$ and thus has no undesirable effect on training.

\begin{figure}[t]
    \scriptsize
    \medskip
    \centering
    \begin{minipage}[b]{0.4\linewidth}
      \centering{3VTV model} \\
      \includegraphics[width=\linewidth]{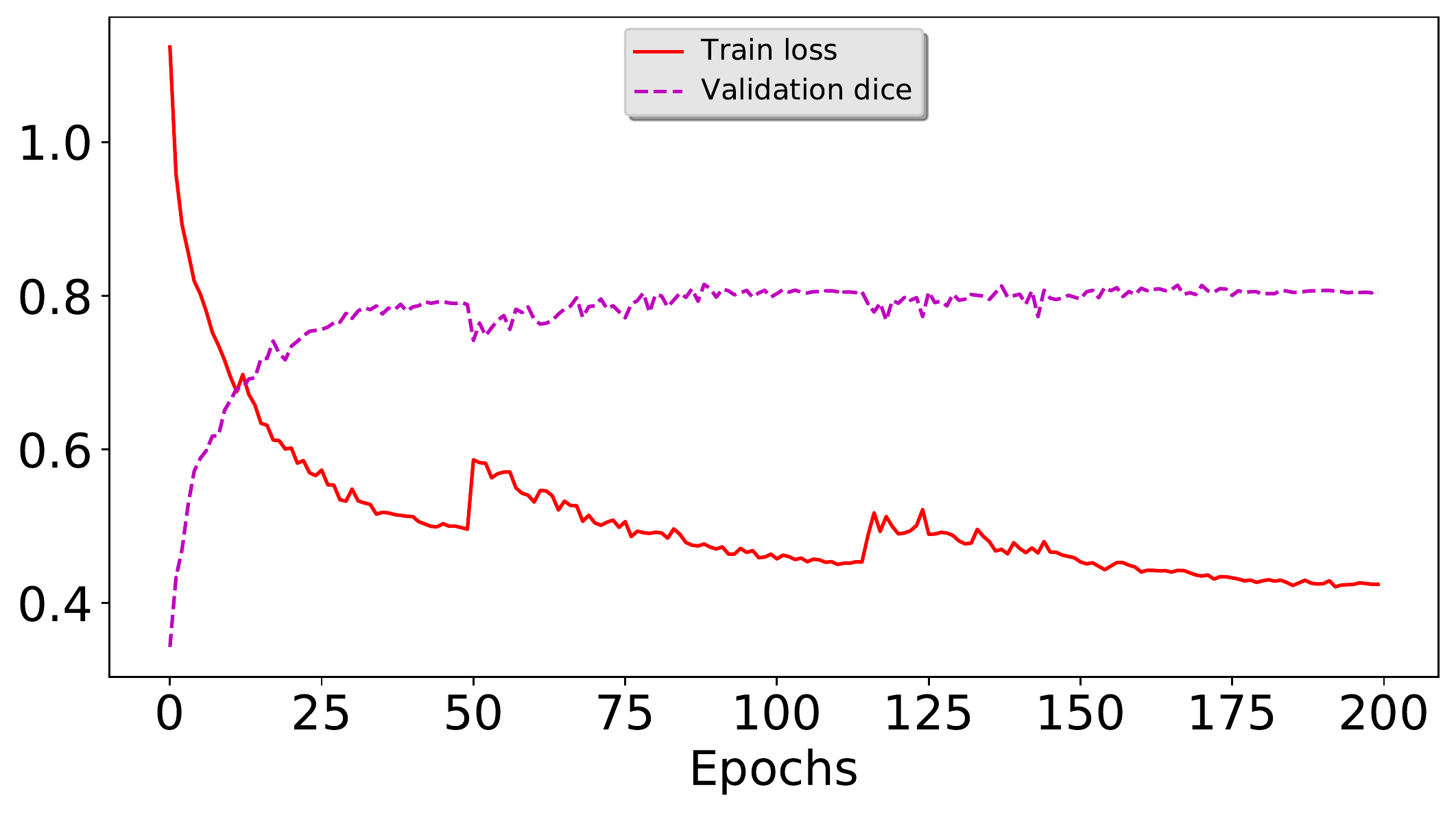}
    \end{minipage}
    \begin{minipage}[b]{0.4\linewidth}
      \centering{4CHV model}
      \includegraphics[width=\linewidth]{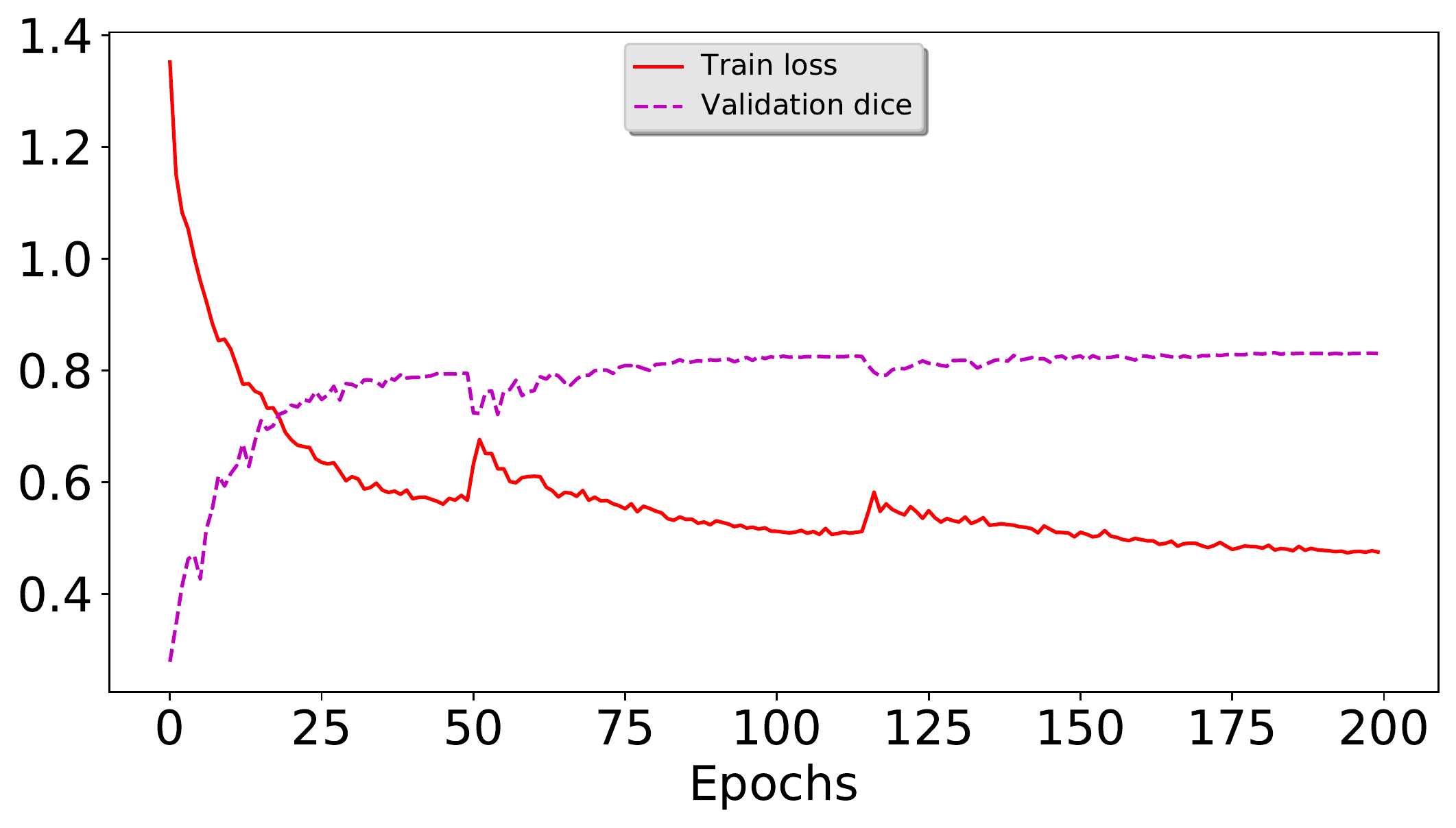}
    \end{minipage}
    \\
    \medskip
    \centering
    \begin{minipage}[b]{0.4\linewidth}
      \centering{Combined-view model}
      \includegraphics[width=\linewidth]{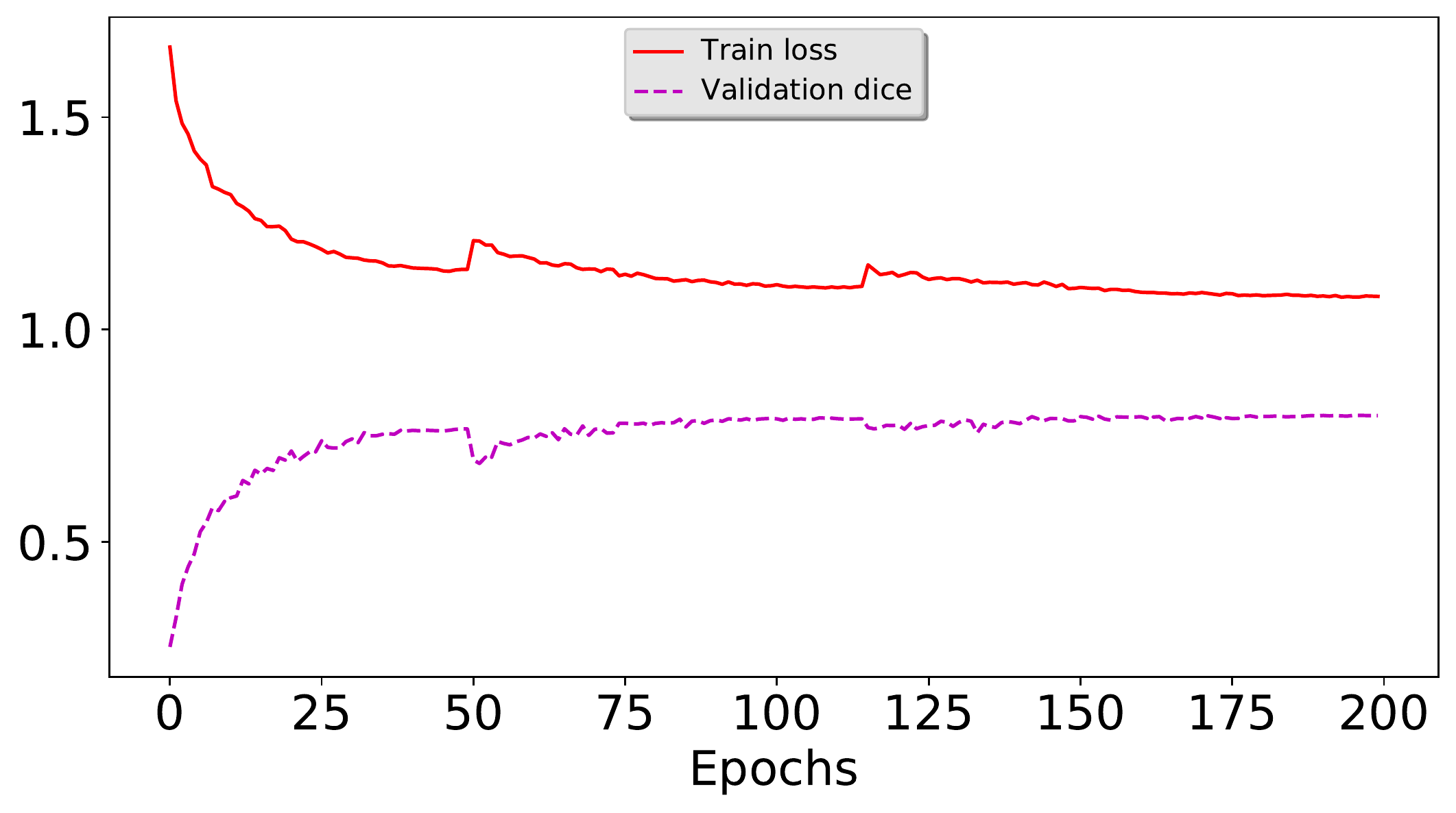}
    \end{minipage}
    \medskip
    \caption{The training loss and validation Dice coefficient \emph{vs}. epoch during training.}
    \label{fig:train_curves}
    \bigskip
\end{figure}

\subsection{Training strategy}

Given the large variations of anatomical sizes and locations, image augmentations with rotation ($\pm30^{\circ}$), shifting ($\pm20\%$), resizing ([0.8, 1.2]), and horizontal flipping were used, and each image had an 80\% chance to be transformed. Image intensity centering was applied on each image. The SGD optimizer with momentum of 0.9 was used. A warm restart learning rate schedule with cosine annealing of three cycles was used \cite{Conference:Loshchilov:ICLR2017:SGDR}, with the minimum and maximum learning rates as $10^{-4}$ and $5\times10^{-3}$, respectively. The scheduler initially restarted at the 50\textsuperscript{th} epoch which was increased by a factor of 1.31 at every restart, with a total of 200 epochs. There was no decay of the learning rate at restarts. The IBM Power System AC922 equipped with NVLink for enhanced host to GPU communication was used. This machine features NVIDIA Tesla V100 GPUs with 16 GB memory, and two of these GPUs were used for multi-GPU training with a batch size of two.

\section{RESULTS}

We split the dataset into training (70\%), validation (10\%), and testing (20\%) portions in terms of patients, and the ratios of the normal and abnormal cases were the same in all portions. Three models were trained using the 3VTV (5 labels), 4CHV (10 labels), and combined-view (14 labels) datasets, respectively.

Fig. \ref{fig:train_curves} shows the curves of training losses and validation Dice coefficients during training. First of all, the curves show the use of the warm restart learning rate schedule with cosine annealing. In each model training, although the learning rate went back to the maximum at each restart, the training loss only increased a relatively small amount, and such increase became smaller in the next restart. This shows the good converging property of our proposed framework. Secondly, the training losses of the single-view models were less than 0.5 at the last epoch, while that of the combined-view model was larger than one. On the other hand, the validation Dice coefficients of all models were similar. Such difference in training losses was caused by the number of unavailable labels in each image as discussed in Section \ref{sec:loss}. When training the combined-view model with all 14 labels from both views, many labels were unavailable for each image. For example, a 3VTV image only had label 6 to 10 as shown in Table \ref{table:data}. The soft Dice coefficients were constantly zeros for these unavailable labels, and this increased the training loss but did not contribute to the learning.

\begin{table*}[t]
\caption{Numerical segmentation results. Testing Dice coefficients between prediction and ground truth averaged from images that have the corresponding labels (format: mean$\pm$std\%). Please refer to Table \ref{table:data} for the semantic labels.}
\label{table:results}

\smallskip
\scriptsize
\centering

\newcolumntype{R}{>{\raggedleft\arraybackslash}X}

\begin{tabularx}{\linewidth}{lRlRlRlRlRlRlRl}
\toprule
\multirow{2}{0.12\linewidth}{3VTV model}
& 1. & --- & 2. & --- & 3. & --- & 4. & --- & 5. & --- & 6. & 82$\pm$13 & 7. & 80$\pm$18 \\
& 8. & 79$\pm$20 & 9. & 67$\pm$20 & 10. & 82$\pm$17 & 11. & --- & 12. & --- & 13. & --- & 14. & --- \\
\midrule
\multirow{2}{0.12\linewidth}{4CHV model}
& 1. & 83$\pm$18 & 2. & 83$\pm$18 & 3. & 85$\pm$12 & 4. & 89$\pm$13 & 5. & 82$\pm$15 & 6. & --- & 7. & --- \\
& 8. & --- & 9. & --- & 10. & 87$\pm$7 & 11. & 81$\pm$10 & 12. & 68$\pm$16 & 13. & 69$\pm$22 & 14. & 74$\pm$16 \\
\midrule
\multirow{2}{0.12\linewidth}{Combined-view model}
& 1. & 83$\pm$18 & 2. & 84$\pm$17 & 3. & 84$\pm$16 & 4. & 90$\pm$11 & 5. & 81$\pm$17 & 6. & 81$\pm$14 & 7. & 80$\pm$16 \\
& 8. & 78$\pm$22 & 9. & 66$\pm$18 & 10. & 86$\pm$8 & 11. & 81$\pm$14 & 12. & 66$\pm$19 & 13. & 69$\pm$22 & 14. & 74$\pm$18 \\
\bottomrule
\end{tabularx}
\end{table*}

\begin{figure}[t]
    \scriptsize
    \medskip
    \centering
    \begin{minipage}[b]{0.49\linewidth}
      \begin{minipage}[b]{0.32\linewidth}
      \centering
      \includegraphics[width=\linewidth]{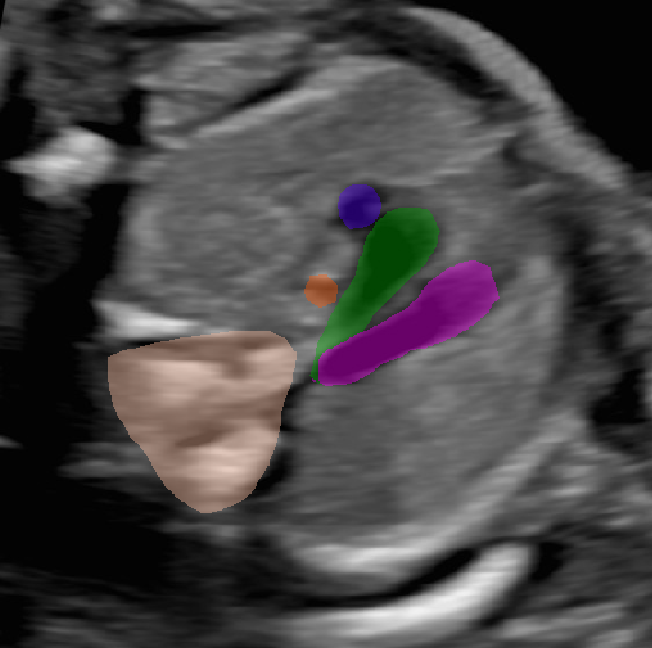}\\
      \centering{Ground truth \phantom{aaaaaaaaaaaaa}}
      \end{minipage}
      \begin{minipage}[b]{0.32\linewidth}
      \centering
      \includegraphics[width=\linewidth]{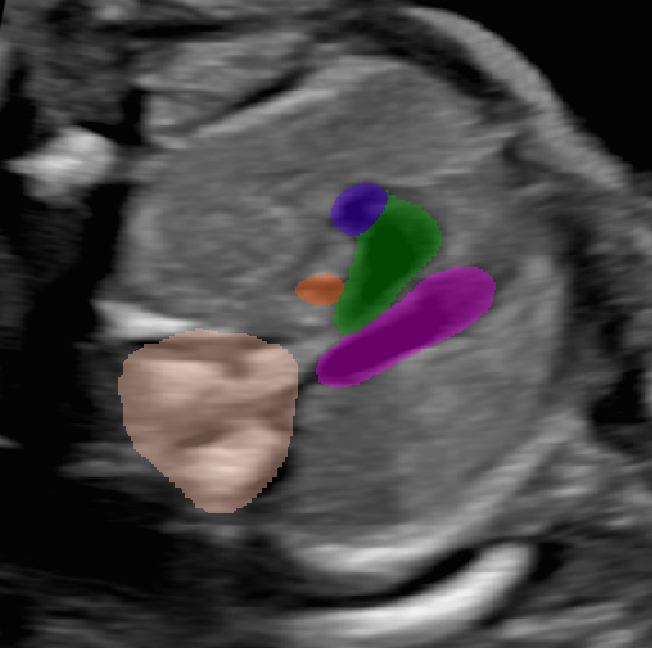} \\
      \centering{3VTV model \phantom{aaaaaaaaaaaaa}}
      \end{minipage}
      \begin{minipage}[b]{0.32\linewidth}
      \centering
      \includegraphics[width=\linewidth]{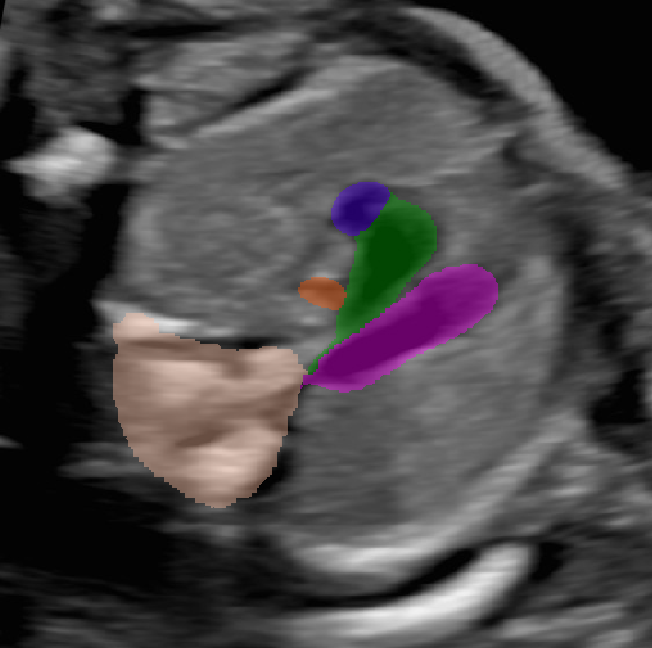} \\
      \centering{Combined-view model}
      \end{minipage}
      \centering Three-vessel trachea view
    \end{minipage}
    \vrule\
    \begin{minipage}[b]{0.49\linewidth}
      \begin{minipage}[b]{0.32\linewidth}
      \centering
      \includegraphics[width=\linewidth]{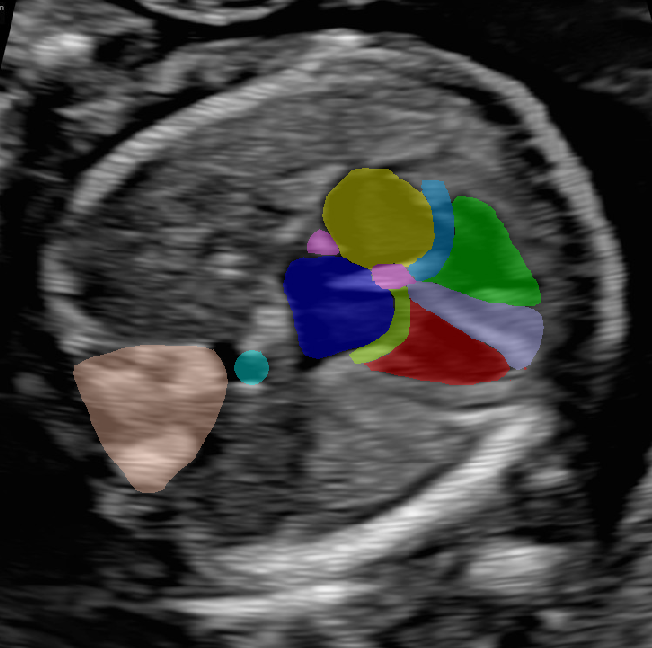}\\
      \centering{Ground truth \phantom{aaaaaaaaaaaaa}}
      \end{minipage}
      \begin{minipage}[b]{0.32\linewidth}
      \centering
      \includegraphics[width=\linewidth]{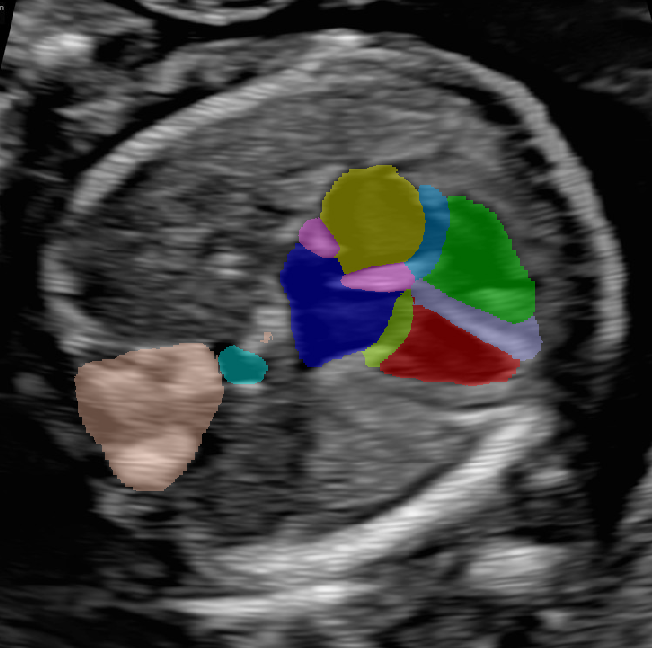} \\
      \centering{4CHV model \phantom{aaaaaaaaaaaaa}}
      \end{minipage}
      \begin{minipage}[b]{0.32\linewidth}
      \centering
      \includegraphics[width=\linewidth]{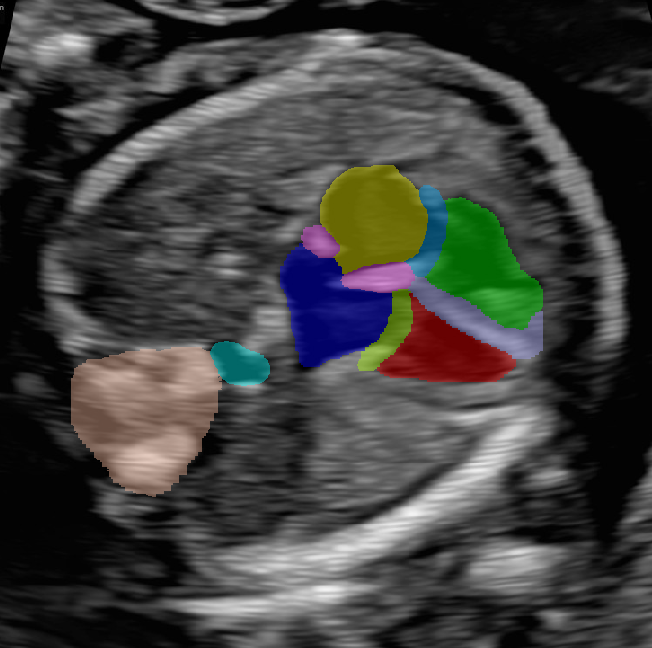} \\
      \centering{Combined-view model}
      \end{minipage}
      \centering Four‐chamber view
    \end{minipage}\\
    \centering(a) A normal case.
    \\
    \medskip
    \begin{minipage}[b]{0.49\linewidth}
      \begin{minipage}[b]{0.32\linewidth}
      \centering
      \includegraphics[width=\linewidth]{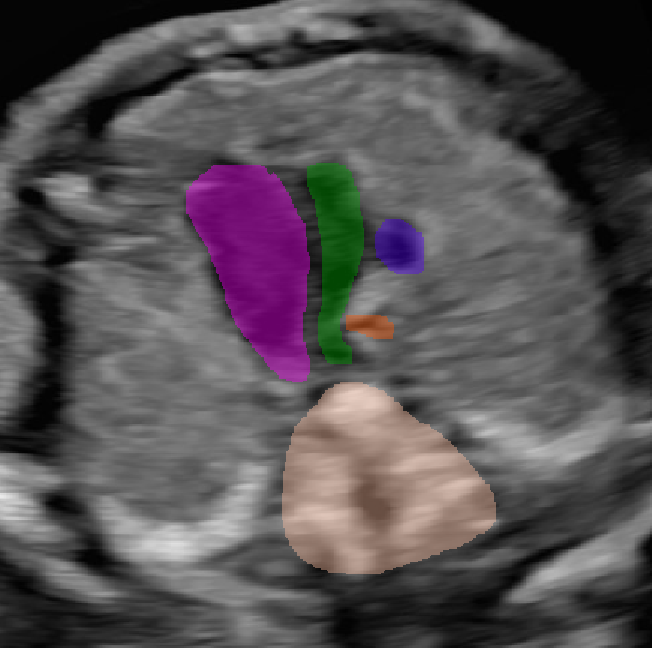}\\
      \centering{Ground truth \phantom{aaaaaaaaaaaaa}}
      \end{minipage}
      \begin{minipage}[b]{0.32\linewidth}
      \centering
      \includegraphics[width=\linewidth]{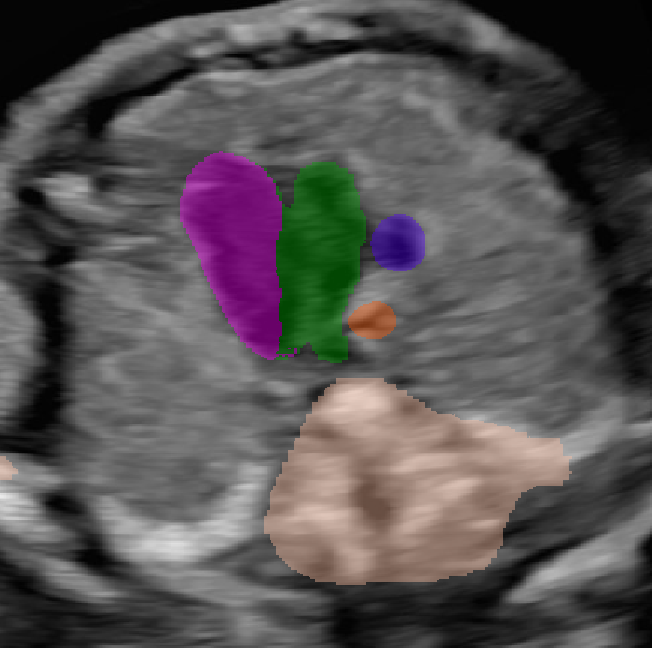} \\
      \centering{3VTV model \phantom{aaaaaaaaaaaaa}}
      \end{minipage}
      \begin{minipage}[b]{0.32\linewidth}
      \centering
      \includegraphics[width=\linewidth]{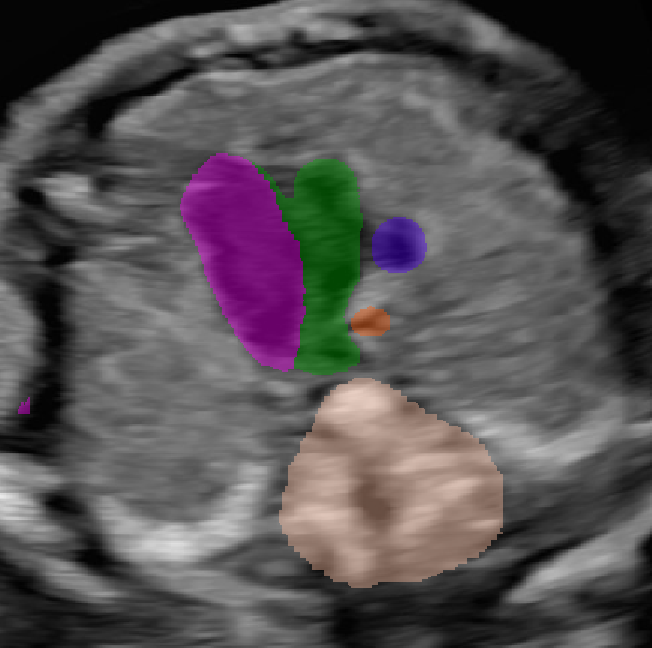} \\
      \centering{Combined-view model}
      \end{minipage}
      \centering Three-vessel trachea view
    \end{minipage}
    \vrule\
    \begin{minipage}[b]{0.49\linewidth}
      \begin{minipage}[b]{0.32\linewidth}
      \centering
      \includegraphics[width=\linewidth]{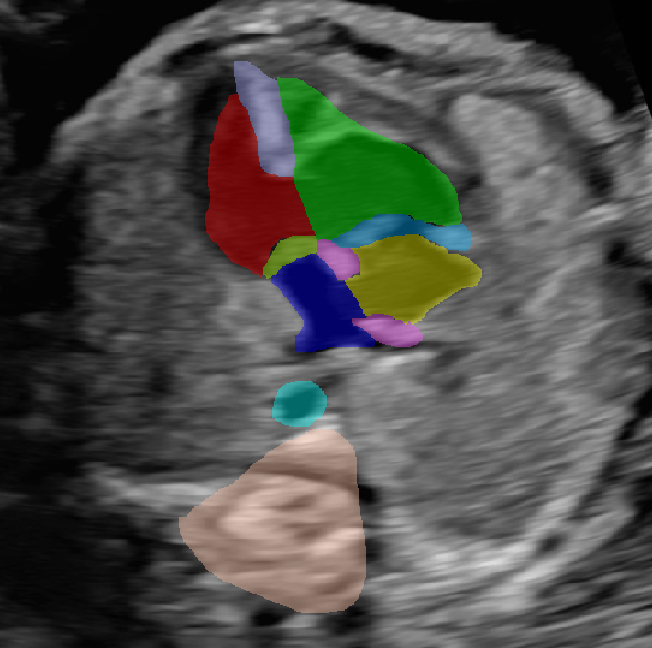}\\
      \centering{Ground truth \phantom{aaaaaaaaaaaaa}}
      \end{minipage}
      \begin{minipage}[b]{0.32\linewidth}
      \centering
      \includegraphics[width=\linewidth]{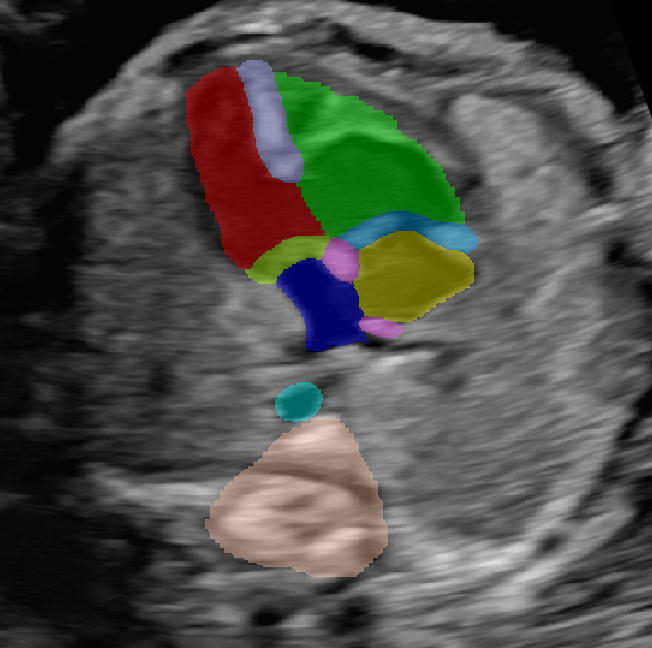} \\
      \centering{4CHV model \phantom{aaaaaaaaaaaaa}}
      \end{minipage}
      \begin{minipage}[b]{0.32\linewidth}
      \centering
      \includegraphics[width=\linewidth]{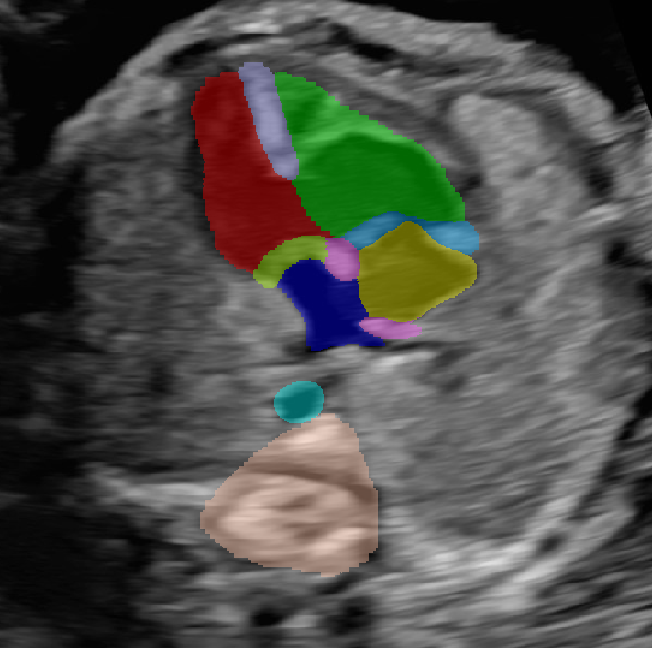} \\
      \centering{Combined-view model}
      \end{minipage}
      \centering Four‐chamber view
    \end{minipage}\\
    \centering(b) An abnormal case.
    \medskip
    \caption{Visualization of segmentation results of normal and abnormal cases in different views.}
    \label{fig:visu}
    \medskip
\end{figure}

Table \ref{table:results} shows the testing Dice coefficients averaged from images that had the corresponding labels. The left and right ventricles and atriums were well segmented (label \#1 -- \#4), with the Dice coefficients larger than 83\% regardless of models. The trachea (label \#9), interatrial septum (label \#12), and mitral valve (label \#13) were more difficult to segment, with the Dice coefficients below 70\%. The standard deviation in each label was large which reflects the large variation in image appearances. Comparing the combined-view model with the 3VTV and 4CHV models, the differences were less than 1\%.

Fig. \ref{fig:visu} shows segmentation examples of using different models in different views. As a fetus can move in the womb, there were large variations in anatomical positions and appearances. Despite this, the segmentation results of all models were very similar to the ground truths. Furthermore, consistent with Table \ref{table:results}, the combined-view model performed as well as the 3VTV and 4CHV models.

\section{CONCLUSION}

We created a fetal echocardiographic dataset that comprises 14 anatomically important structures for interpreting congenital anomalies. By improving the V-Net with spatial dropout, group normalization, and deep supervision, and by identifying the pitfall of using the Dice loss when some labels are unavailable, we can train a model to accurately segment fetal echocardiograms in multiple views and abnormal conditions. Comparing the combined-view model with the 3VTV and 4CHV models, there was no reduction in segmentation performance. This shows that the proposed framework is adaptive to large variations in appearances and annotations, and multiple models for multiple views are unnecessary.

\bibliography{Ref}
\bibliographystyle{spiebib}   

\end{document}